\documentclass[aps,prl,twocolumn,superscriptaddress,floatfix,10pt]{revtex4-2}

\usepackage{amsmath,amssymb,amsfonts}
\usepackage{graphicx}
\usepackage{bm} 
\usepackage[svgnames]{xcolor}
\usepackage{siunitx} 
\usepackage{microtype} 
\usepackage{balance} 
\usepackage{hyperref} 

\graphicspath{{plots1/}}

\hypersetup{
    colorlinks=true,
    linkcolor=NavyBlue,
    citecolor=ForestGreen,
    urlcolor=SteelBlue,
    pdfauthor={Shivanshu Dwivedi and Kalum Palandage},
    pdftitle={Geometric and Orbital Control of Correlated States in Small Hubbard Clusters}
}

\newcommand{\crea}[3]{\hat{c}_{#1,#2,#3}^{\dagger}}
\newcommand{\anni}[3]{\hat{c}_{#1,#2,#3}}
\newcommand{\nhat}[3]{\hat{n}_{#1,#2,#3}} 
\newcommand{\expect}[1]{\langle #1 \rangle}
\newcommand{\torb}{t_{\mathrm{orb}}}

\begin{document}

\title{Geometric and Orbital Control of Correlated States in Small Hubbard Clusters}

\author{Shivanshu Dwivedi}
\author{Kalum Palandage}
\affiliation{Department of Physics, Trinity College, Hartford, CT 06106, USA}

\date{\today}

\begin{abstract}
Arrays of semiconductor quantum dots provide a powerful platform to design correlated quantum matter from the bottom up. We establish a predictive framework for engineering local electron pairing in these artificial molecules by systematically deploying three control levers: lattice geometry, orbital hybridization, and external electric fields. Using Hartree-Fock simulations on canonical 3D clusters from the tetrahedron ($Z=3$) to the FCC lattice ($Z=12$), at and near half-filling, we uncover three fundamental design principles. (i) \textbf{Geometric Hierarchy:} The resilience to Coulomb repulsion $U$ is dictated by the coordination number $Z$, which controls kinetic delocalization. (ii) \textbf{Orbital Hybridization:} Counter-intuitively, inter-orbital hopping $\torb$ acts not as a simple suppressor of pairing, but as a sophisticated control knob that \emph{enhances} double occupancy at moderate $U$ by engineering the on-site energy landscape. (iii) \textbf{Field Squeezing:} An electric field robustly induces pairing by forcing charge localization, an effect most potent in low-connectivity clusters. These principles form a blueprint for deterministically targeting charge and spin correlations in quantum-dot-based quantum hardware.
\end{abstract}

\maketitle


\section{Introduction}

Arrays of semiconductor quantum dots offer a tangible realization of the Hubbard model~\cite{Hubbard1963, Gutzwiller1963}, providing an ideal platform to engineer correlated quantum matter from the bottom up~\cite{Hensgens2017, Bloch2008, Esslinger2010}. As this technology matures into scalable quantum hardware, a central challenge emerges: establishing the design rules to deterministically control the system's many-body electronic states~\cite{Loss1998, Imada1998, Lee2006}. The key lies in manipulating local electron pairing ~\cite{Kocharian2005JMMM, Kocharian2006PRB, Kocharian2007PLA, Kocharian2008PRB, Kocharian2009PLA, Fernando2007PRB, Fernando2009PRB, Palandage2007JCAMD, Tsai2006}. The double occupancy is not merely a diagnostic, but a direct measure of the charge and spin correlations that govern the system. Control over this quantity is tantamount to engineering the exchange interactions and energy splittings essential for qubit operations and for tuning the system across distinct phases of matter~\cite{Loss1998, Georges1996, Kotliar2006}.

In this paper, we provide these design rules by systematically decoupling and quantifying the three fundamental levers of control: lattice geometry, on-site orbital hybridization, and external electric fields. We analyze a spectrum of canonical 3D clusters—from the tetrahedron ($Z=3$) to the FCC lattice ($Z=12$)—at and near half-filling, where correlation effects are most pronounced. Our results uncover a set of universal principles for quantum state engineering, including the counter-intuitive role of inter-orbital hopping as a promoter of pairing at moderate interaction strengths. Together, these principles form a predictive blueprint for designing quantum dot arrays with precisely targeted functionalities.

\section{Theoretical Framework}

\subsection{The Multi-Orbital Hubbard Hamiltonian}

Our system of interacting electrons in a finite quantum dot array is described by the multi-orbital Hubbard Hamiltonian~\cite{Hubbard1963}. In its general form, it captures kinetic hopping between any site and orbital combination, and an orbital-dependent on-site repulsion:
\begin{widetext}
\begin{equation}
\hat{H} = - \sum_{\langle i,j \rangle, \alpha, \beta, \sigma} 
t_{\alpha\beta}^{ij} \left( \crea{i}{\alpha}{\sigma} \anni{j}{\beta}{\sigma} + \text{h.c.} \right)
+ \sum_{i, \alpha} U_{\alpha} \, \nhat{i}{\alpha}{\uparrow} \, \nhat{i}{\alpha}{\downarrow}
- \sum_{i,\alpha,\sigma} \left(e\mathbf{r}_i \cdot \mathbf{E} + \mu \right) \nhat{i}{\alpha}{\sigma}.
\label{eq:hubbard_general}
\end{equation}
\end{widetext}

Here, $\crea{i}{\alpha}{\sigma}$ creates an electron of spin $\sigma$ in orbital $\alpha$ at site $i$. The hopping matrix $t_{\alpha\beta}^{ij}$ describes the amplitude for an electron to move from orbital $\beta$ at site $j$ to orbital $\alpha$ at site $i$, and $U_\alpha$ is the energy cost for double occupancy of orbital $\alpha$.

For this study, we focus on a specific, physically motivated realization of this model that isolates the key control levers. We make the following assumptions: (i) hopping is restricted to nearest-neighbor sites ($i, j$ are adjacent) or occurs on the same site ($i=j$), (ii) the on-site repulsion $U$ is uniform for all orbitals, and (iii) inter-site hopping does not change the orbital index. This simplifies the general Hamiltonian into distinct terms whose physical roles can be systematically investigated:

\begin{enumerate}
    \item \textbf{Inter-site hopping ($t_0$):} This term, corresponding to the $t_{\alpha\alpha}^{ij}$ elements for nearest neighbors, governs electron tunneling between adjacent quantum dots. We recognize that in realistic finite clusters, not all nearest-neighbor bonds are equivalent, leading to a set of distinct hopping parameters (e.g., $t_{\text{base}}$, $t_{\text{vertical}}$) dictated by the geometry. To establish a single, effective hopping parameter that faithfully represents the overall kinetic character of each unique topology, we employ a weighted-average optimization. This procedure involves defining an effective $t_0$ from its constituent geometric components and then maximizing a measure of electron delocalization to find the most representative value. This optimized, effective $t_0$ then serves as the fundamental unit of energy ($t_0 \equiv 1$) for all subsequent calculations, ensuring a physically meaningful and consistent basis for comparing the diverse geometries in our study.
    
    \item \textbf{Intra-site hopping ($\torb$):} This corresponds to $t_{\alpha\beta}^{ii} = \torb$ for $\alpha \neq \beta$. It represents quantum tunneling between different orbitals on the \emph{same} site. This term, controllable via gate voltages or strain, provides a local, intra-dot delocalization pathway that directly competes with the formation of a doubly-occupied state in a single orbital.
    
    \item \textbf{On-site repulsion ($U$):} We set $U_\alpha = U$ for all orbitals. This is the standard Hubbard interaction, which penalizes the double occupancy of any given orbital with an energy cost $U$.
    
    \item \textbf{Local Potentials:} The final term includes the potential energy from a static external electric field $\mathbf{E}$ at each site position $\mathbf{r}_i$, and the chemical potential $\mu$, which is adjusted to fix the total number of electrons in the cluster, thereby controlling the filling.
\end{enumerate}

We analyze this model at half-filling ($\langle \hat{N} \rangle = N_{\text{sites}} \times N_{\text{orbitals}}$) and in the doped regime, one electron removed from half-filling.

\begin{figure*}[t!]
    \centering
    \includegraphics[width=\textwidth]{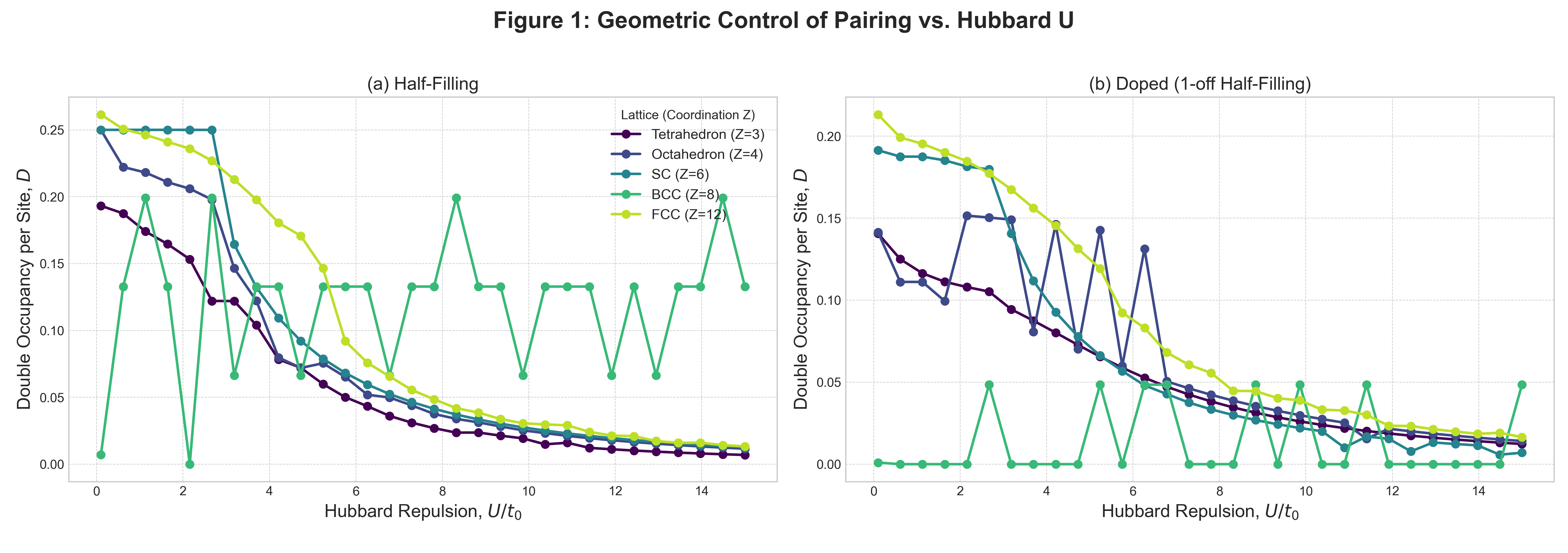}
    \caption{\textbf{Geometric control of pairing vs. Hubbard U.} Total double occupancy ($D$) as a function of Hubbard repulsion ($U/t_0$) for five different lattice geometries. (a) At half-filling, a clear hierarchy emerges where higher coordination ($Z$) sustains greater pairing against $U$. (b) In the doped case (one electron removed), the hierarchy is less rigid, and the suppression of $D$ is more gradual.}
    \label{fig:1}
\end{figure*}

\subsection{Computational Methods}

To solve the many-body problem posed by Eq.~(\ref{eq:hubbard_general}), we employ a dual approach tailored to the system size. For small clusters where the Hilbert space is manageable (the 4-site tetrahedron and 6-site octahedron), we use \textbf{exact diagonalization} (ED)~\cite{Kocharian2006PRB, Kocharian2007PLA, Kocharian2008PRB}. This numerically exact, unbiased method provides a crucial benchmark for the ground state and thermodynamic properties, capturing all quantum correlations without approximation.

For a systematic survey across all five geometries, including the larger SC, BCC, and FCC clusters, and for the broad parameter sweeps required to map out the phase space, we use a computationally efficient \textbf{self-consistent Hartree-Fock} (HF) method. This mean-field approach decouples the four-fermion interaction term into an effective potential determined by the average site and orbital occupations:
\begin{equation}
U \hat{n}_{\uparrow} \hat{n}_{\downarrow} \approx U \left(\expect{\hat{n}_{\uparrow}}\hat{n}_{\downarrow} + \hat{n}_{\uparrow}\expect{\hat{n}_{\downarrow}} - \expect{\hat{n}_{\uparrow}}\expect{\hat{n}_{\downarrow}}\right).
\end{equation}
The resulting quadratic Hamiltonian is diagonalized, and the new occupation numbers $\expect{\hat{n}_{i\alpha\sigma}}$ are calculated from the resulting eigenstates. These are then fed back into the effective potential, and the process is iterated until the occupations converge to a self-consistent solution. While this method neglects quantum fluctuations beyond the mean-field level, it effectively captures the essential physics of local correlation and charge ordering~\cite{Tsai2006}, making it ideal for a comprehensive study of the parameter space.

A key challenge in comparing topologically distinct lattices is establishing a consistent energy scale. As described in the previous section, the inter-site hopping $t_0$ can have multiple inequivalent values within a single cluster. To define a single, effective $t_0$ that robustly characterizes the kinetic properties of each geometry, we analyze the single-particle spectrum of the non-interacting Hamiltonian ($U=0$). We define an effective hopping $t_0$ as a function of its geometric components (e.g., $t_{\text{base}}$, $t_{\text{vertical}}$) and numerically find the parameter combination that maximizes the kinetic bandwidth $W$—the energy difference between the highest and lowest single-particle eigenvalues. This procedure identifies the hopping configuration that maximizes electron delocalization for a given topology. This optimized, effective hopping is then set as the unit of energy ($t_0 \equiv 1$) for that geometry, ensuring that our cross-topological comparisons are made on a consistent and physically meaningful basis.

Our primary observable is the average double occupancy per site, $D$, which directly quantifies the degree of local electron pairing:
\begin{equation}
D = \frac{1}{N_{\text{sites}}}\sum_{i,\alpha} \expect{\nhat{i}{\alpha}{\uparrow} \nhat{i}{\alpha}{\downarrow}}.
\end{equation}

\begin{figure*}[t!]
    \centering
    \includegraphics[width=0.98\textwidth]{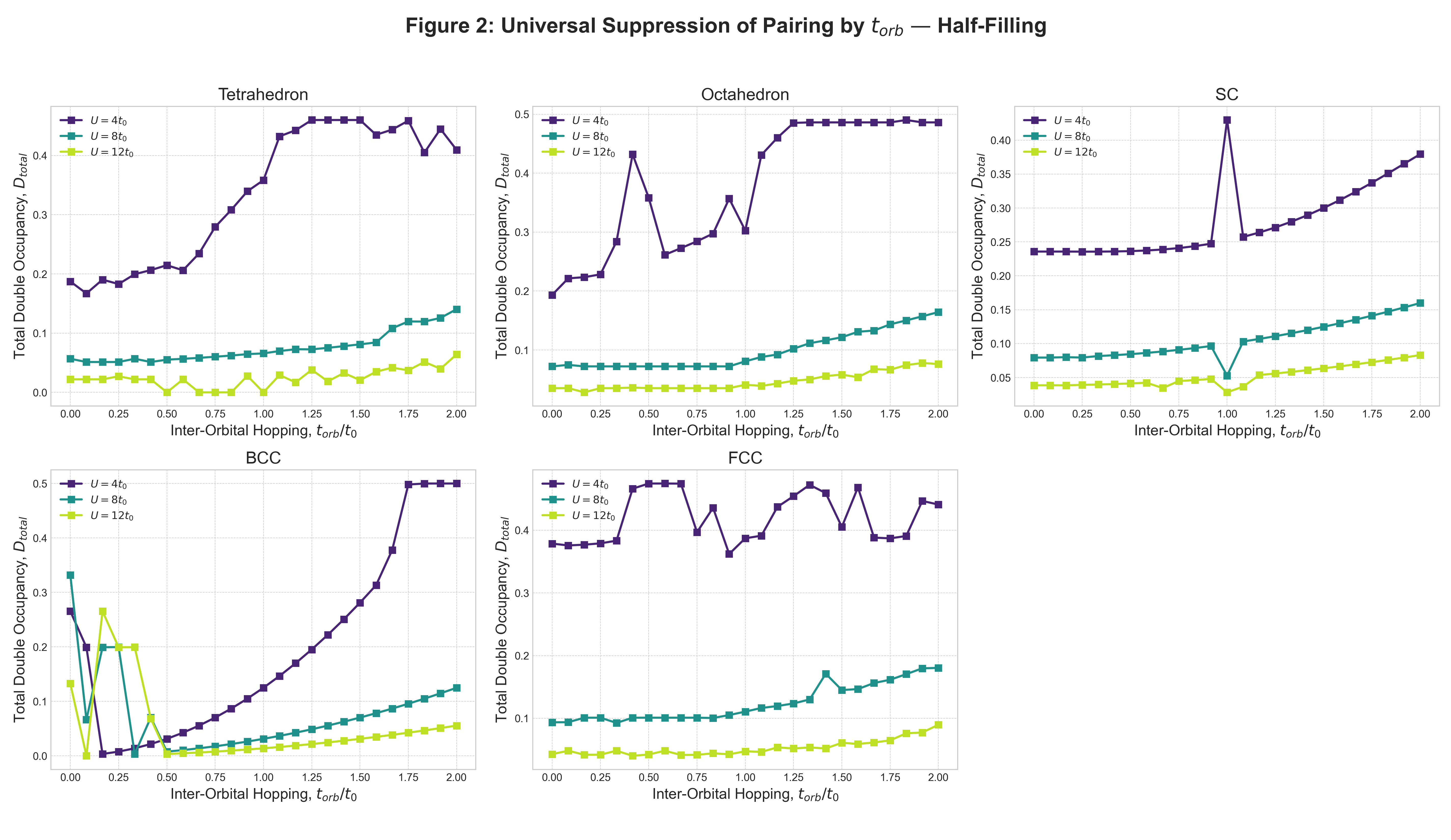}
    \vspace{0.5cm} 
    \includegraphics[width=0.98\textwidth]{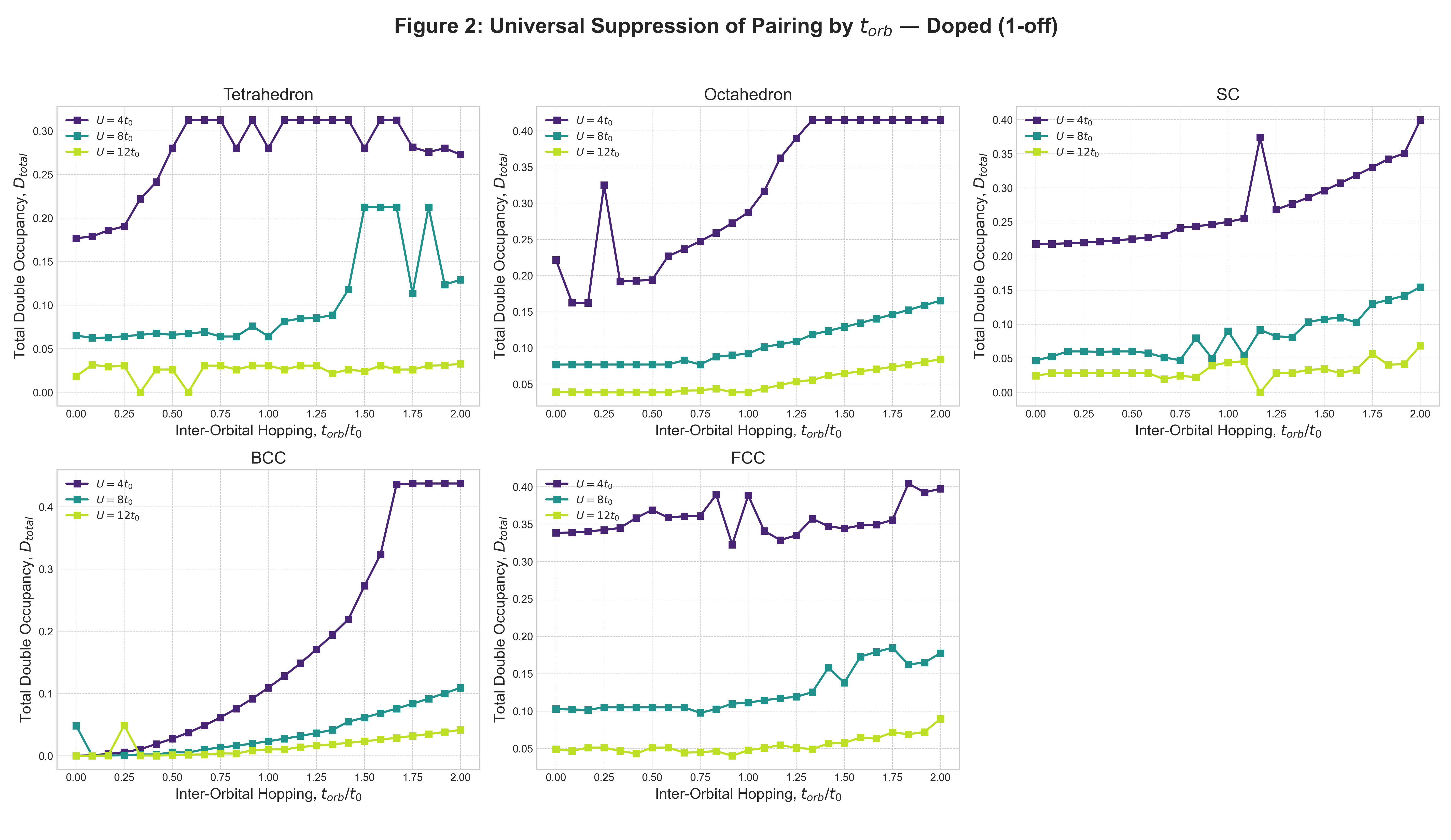}
    \caption{\textbf{Control of pairing via inter-orbital hopping $\torb$.} Total double occupancy ($D_{\text{total}}$) versus $\torb/t_0$ for all geometries at weak ($U=4t_0$), intermediate ($U=8t_0$), and strong ($U=12t_0$) coupling. \textbf{(Top Panel)} At half-filling, $\torb$ strongly enhances pairing at moderate $U$, while its effect is quenched at large $U$. \textbf{(Bottom Panel)} In the doped case, the enhancement is more modest but follows a similar trend.}
    \label{fig:2}
\end{figure*}

\section{Results and Discussion}

\subsection{Intrinsic Control I: Geometric Hierarchy}

\begin{figure*}[t!]
    \centering
    \includegraphics[width=\textwidth]{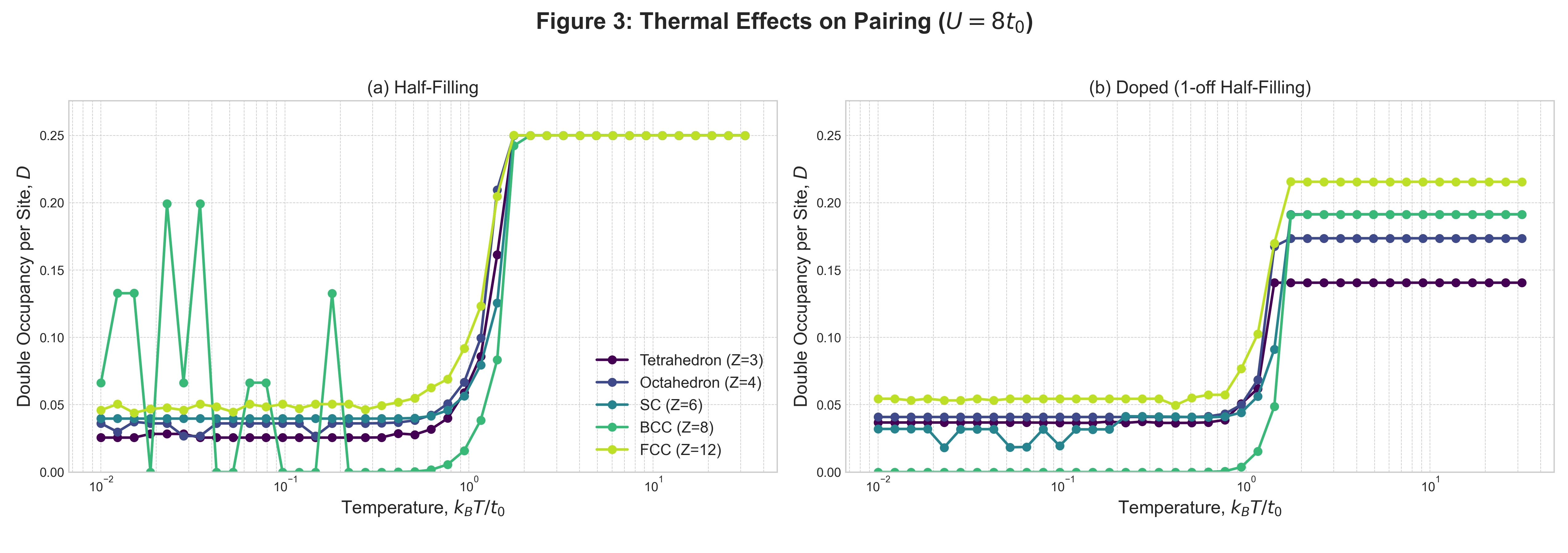}
    \caption{\textbf{Thermal effects on pairing at strong coupling ($U=8t_0$).} Double occupancy ($D$) versus temperature. (a) At half-filling, all geometries exhibit a sharp thermal crossover from a low-pairing Mott state to a high-temperature disordered state, universally saturating at $D=0.25$. (b) In the doped case, the system saturates to a lower, geometry-dependent value of $D$, reflecting the reduced particle density.}
    \label{fig:3}
\end{figure*}

The primary intrinsic control parameter for electron pairing is the lattice geometry itself, quantified by the coordination number $Z$. Figure~\ref{fig:1} reveals how this topological feature governs the system's response to Coulomb repulsion, a behavior that is starkly dependent on electron filling.

At half-filling [Fig.~\ref{fig:1}(a)], a robust geometric hierarchy emerges: higher connectivity grants greater resilience to the on-site repulsion $U$. The enhanced kinetic delocalization in high-$Z$ lattices (e.g., FCC) allows electrons to effectively screen their interaction, thus preserving a higher double occupancy $D$. Conversely, the limited hopping pathways in low-$Z$ clusters (e.g., tetrahedron) cause them to be readily driven into a Mott insulating state~\cite{Kocharian2005JMMM}, evidenced by the rapid quenching of $D$. The non-monotonic fluctuations seen in the SC and BCC clusters are a mean-field signature of level crossings between competing, nearly-degenerate charge and spin configurations inherent to their complex energy landscapes.

Doping the system by one hole fundamentally alters this picture [Fig.~\ref{fig:1}(b)]. The Mott physics vanishes, replaced by the behavior of a correlated metal. The suppression of $D$ with $U$ becomes more gradual across all geometries, as the mobile hole provides an efficient screening channel that forestalls strong localization. While the general influence of $Z$ persists, the rigid hierarchy is softened, and the system's response becomes more sensitive to the specific non-interacting band structure of each geometry. Thus, lattice connectivity is a powerful lever, most effective at half-filling, for tuning the proximity to a Mott transition.

\subsection{Intrinsic Control II: Orbital Hybridization}

The on-site orbital structure provides a second, powerful intrinsic control knob. Tuning the inter-orbital hopping, $\torb$, reveals a profound and counter-intuitive effect that is strongly dependent on the correlation strength $U$, as shown in Fig.~\ref{fig:2}.

Our central finding is that at moderate coupling ($U=4t_0$), $\torb$ acts not to suppress pairing, but to substantially \emph{enhance} it. This phenomenon arises from on-site orbital hybridization. A non-zero $\torb$ creates local bonding and anti-bonding-like states on each site. The kinetic energy gained by two electrons occupying a low-energy bonding state can partially overcome the Coulomb penalty, making double occupancy more energetically favorable. This pairing enhancement is particularly dramatic in the half-filled tetrahedron, octahedron, and BCC lattices.

This mechanism is completely quenched at strong coupling ($U \ge 8t_0$). Here, the Coulomb blockade is the dominant energy scale, and the kinetic gain from hybridization is insufficient to overcome it. As a result, the double occupancy remains suppressed near zero and is largely insensitive to $\torb$. The unique behavior of the BCC lattice at $U=8t_0$ signals a sharp crossover between competing ground states, a feature of its specific topology.

In the doped case [bottom panel of Fig.~\ref{fig:2}], the hybridization-driven pairing enhancement persists but is less pronounced. The presence of mobile holes provides an alternative, more efficient channel for kinetic energy minimization via inter-site transport, thus reducing the relative impact of the on-site mechanism. Thus, $\torb$ is not a simple pairing suppressor but a sophisticated tuning parameter that engineers the on-site energy landscape to either promote or ignore pairing depending on the strength of correlations.

\begin{figure*}[t!]
    \centering
    \includegraphics[width=\textwidth]{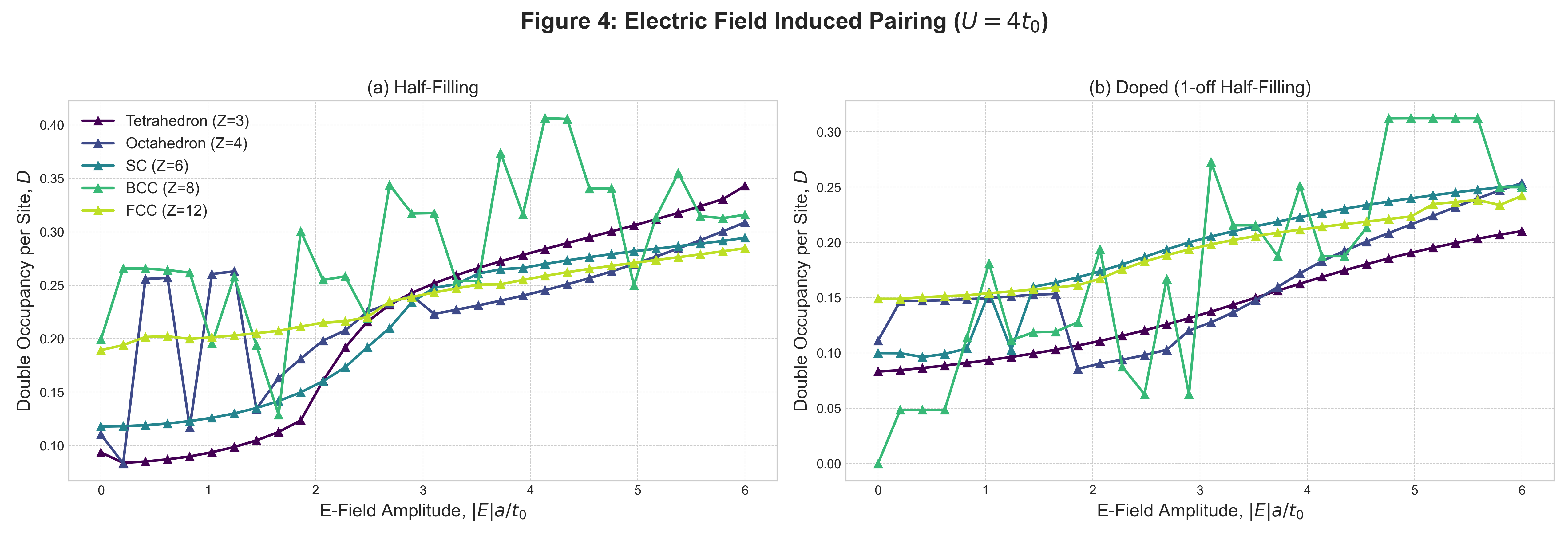}
    \caption{\textbf{Electric field induced pairing at moderate coupling ($U=4t_0$).} Double occupancy ($D$) as a function of E-field amplitude. (a) At half-filling, the field robustly enhances pairing. The effect is dramatically more pronounced in low-coordination lattices (Tetrahedron, Octahedron), which act as effective "quantum squeezers." (b) In the doped case, the overall enhancement is more modest as mobile holes provide alternative charge rearrangement pathways.}
    \label{fig:4}
\end{figure*}

\subsection{Extrinsic Control I: Thermal Effects}

Temperature acts as an extrinsic probe, revealing the characteristic energy scales that govern the correlated system. We investigate its effect at strong coupling ($U=8t_0$), with the results shown in Fig.~\ref{fig:3}.

At half-filling [Fig.~\ref{fig:3}(a)], the system transitions between two clear limits. At low temperatures, it is a Mott insulator with strongly suppressed double occupancy ($D \approx 0$), a direct consequence of the large correlation gap. As $k_B T$ approaches $U$, thermal fluctuations overcome this gap, driving a sharp crossover to a disordered state. In the high-temperature limit, entropy dominates, washing out all quantum correlations and universally driving the system to its classical statistical limit, $D = 0.25$.

Doping fundamentally alters the low-temperature state to a correlated metal, yet the high-temperature physics follows the same entropic principle [Fig.~\ref{fig:3}(b)]. The system again saturates at high T, but to a lower, geometry-dependent value. This reflects the reduced particle density per state, $p < 0.5$. In the high-T limit, the double occupancy approaches $D \approx p^2$, a value unique to the number of available states in each distinct cluster, thereby explaining the clear hierarchy in the saturation plateaus.

\begin{figure*}[t!]
    \centering
    \includegraphics[width=\textwidth]{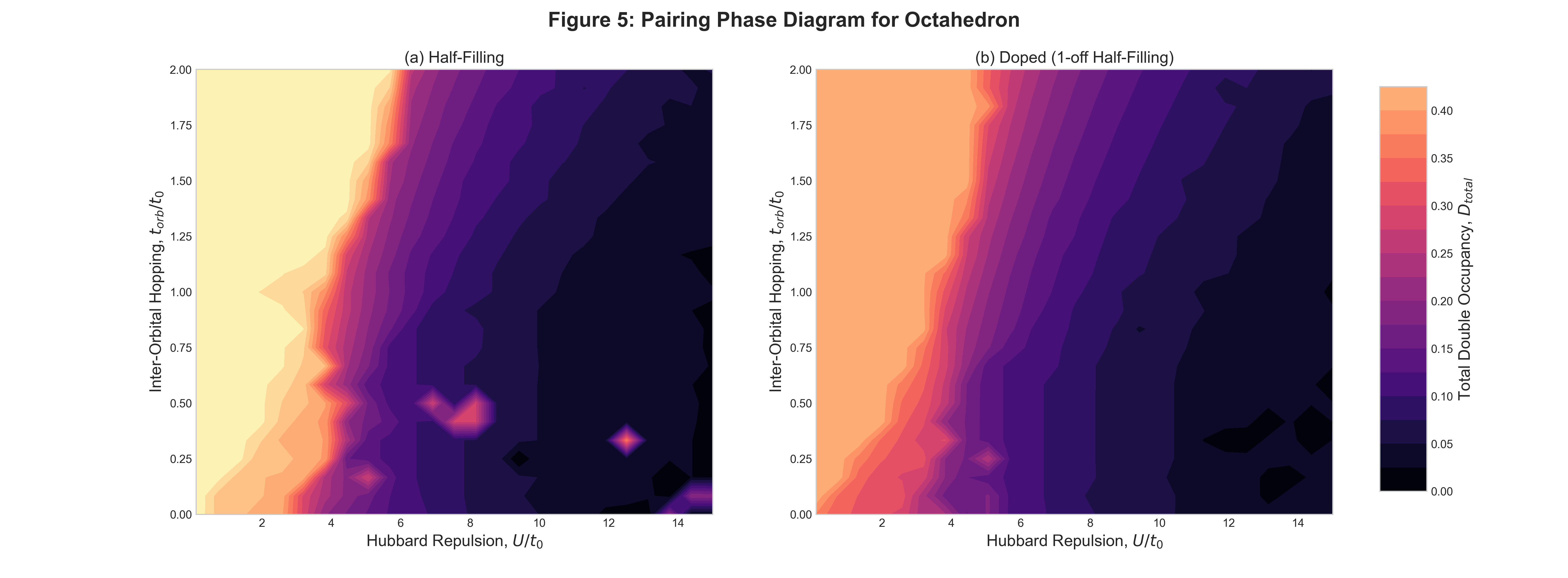}
    \caption{\textbf{Pairing phase diagram for the Octahedron.} Total double occupancy ($D_{\text{total}}$) as a function of Hubbard repulsion $U$ and inter-orbital hopping $\torb$. (a) At half-filling, a clear transition from a high-pairing (bright) to a low-pairing (dark) Mott state is driven by $U$. (b) The transition is smoother in the doped case, consistent with more metallic behavior.}
    \label{fig:5}
\end{figure*}

\subsection{Extrinsic Control II: Electric Field Induced Pairing}

An external electric field acts as a "quantum squeezer," universally enhancing pairing by forcing charge localization, as shown in Fig.~\ref{fig:4}. The efficacy of this mechanism, however, is a direct probe of the competition between the localizing field and the global kinetic energy of the lattice.

At half-filling [Fig.~\ref{fig:4}(a)], the field's potential gradient forces electrons onto low-potential sites, increasing $D$. Crucially, the system's susceptibility to this squeezing is inversely related to its connectivity. Low-$Z$ clusters like the tetrahedron, with their small kinetic bandwidth, are easily polarized, showing a dramatic pairing enhancement. In contrast, the large kinetic energy of high-$Z$ lattices strongly resists this localization, leading to a much weaker response.

This effect is less pronounced in the doped regime [Fig.~\ref{fig:4}(b)]. The presence of mobile holes provides a more efficient channel for charge rearrangement; the system can lower its energy by moving holes to high-potential sites rather than exclusively by forming new pairs. This alternative screening mechanism results in a more gradual increase in $D$. The prominent oscillations seen in the SC and BCC lattices are mean-field signatures of sharp level crossings between competing charge configurations as the field is tuned.

\subsection{Pairing Phase Diagram}

The interplay between the intrinsic control parameters is synthesized in the pairing phase diagram for the octahedron cluster~\cite{Fernando2009PRB}, shown in Fig.~\ref{fig:5}. This map provides an operational blueprint for engineering a target correlation strength, building on prior work on smaller clusters~\cite{Palandage2007JCAMD}.

The diagram is dominated by the Hubbard repulsion: moving horizontally, increasing $U$ drives the system from a weakly correlated state with high pairing (bright) to a strongly correlated, low-pairing state (dark). The vertical axis reveals the more subtle role of $\torb$. Consistent with our earlier findings, it enhances pairing at weak coupling ($U/t_0 \lesssim 4$) via orbital hybridization, but its influence is completely quenched in the strong-coupling, Mott-like regime where $U$ dominates.

The nature of the transition is filling-dependent. The crossover to the low-pairing state is sharp and well-defined at half-filling [Fig.~\ref{fig:5}(a)], a hallmark of the Mott gap. In contrast, the transition is smoother in the doped, metallic case [Fig.~\ref{fig:5}(b)], providing a clear visual distinction between the two regimes.

\section{Conclusion}

In summary, we have established a predictive framework for engineering electron correlations in quantum dot arrays by decoupling three fundamental control levers. Our findings reveal a clear set of design principles: (i) Lattice connectivity ($Z$) dictates the system's resilience to Coulomb repulsion, tuning its proximity to a Mott state. (ii) Counter-intuitively, on-site orbital hybridization ($\torb$) is not a simple pairing suppressor but a sophisticated knob that enhances pairing at moderate interaction strengths. (iii) External electric fields act as "quantum squeezers," robustly inducing pairing with an efficacy inversely related to the lattice connectivity. These principles provide an operational roadmap for the bottom-up design of quantum hardware, demonstrating how the strategic selection of geometry, the engineering of orbital levels, and the application of local fields can be used to navigate the Hubbard parameter space and realize quantum devices with precisely targeted charge and spin functionalities.

\vspace{0.5em}
\balance 

\bibliographystyle{apsrev4-2}

\begin{thebibliography}{19}

\bibitem{Loss1998}
D.~Loss and D.~P.~DiVincenzo,
\textit{Quantum computation with quantum dots},
Phys.\ Rev.\ A \textbf{57}, 120 (1998).
\bibitem{Hubbard1963}
J.~Hubbard,
\textit{Electron Correlations in Narrow Energy Bands},
Proc.\ R.\ Soc.\ Lond.\ A \textbf{276}, 238 (1963).
\bibitem{Gutzwiller1963}
M.~C.~Gutzwiller,
\textit{Effect of Correlation on the Ferromagnetism of Transition Metals},
Phys.\ Rev.\ Lett.\ \textbf{10}, 159 (1963).
\bibitem{Imada1998}
M.~Imada, A.~Fujimori, and Y.~Tokura,
\textit{Metal-insulator transitions},
Rev.\ Mod.\ Phys.\ \textbf{70}, 1039 (1998).
\bibitem{Lee2006}
P.~A.~Lee, N.~Nagaosa, and X.-G.~Wen,
\textit{Doping a Mott insulator: Physics of high-temperature superconductivity},
Rev.\ Mod.\ Phys.\ \textbf{78}, 17 (2006).
\bibitem{Hensgens2017}
T.~Hensgens, T.~Fujita, L.~Janssen, X.~Li, C.~J.~Van Diepen, C.~Reichl, W.~Wegscheider, S.~Das~Sarma, and L.~M.~K.~Vandersypen,
\textit{Quantum simulation of a Fermi-Hubbard model using a semiconductor quantum dot array},
Nature \textbf{548}, 70 (2017).
\bibitem{Bloch2008}
I.~Bloch, J.~Dalibard, and W.~Zwerger,
\textit{Many-body physics with ultracold gases},
Rev.\ Mod.\ Phys.\ \textbf{80}, 885 (2008).
\bibitem{Esslinger2010}
T.~Esslinger,
\textit{Fermi-Hubbard Physics with Atoms in an Optical Lattice},
Annu.\ Rev.\ Condens.\ Matter Phys.\ \textbf{1}, 129 (2010).
\bibitem{Georges1996}
A.~Georges, G.~Kotliar, W.~Krauth, and M.~J.~Rozenberg,
\textit{Dynamical mean-field theory of strongly correlated fermion systems and the limit of infinite dimensions},
Rev.\ Mod.\ Phys.\ \textbf{68}, 13 (1996).
\bibitem{Kotliar2006}
G.~Kotliar, S.~Y.~Savrasov, K.~Haule, V.~S.~Oudovenko, O.~Parcollet, and C.~A.~Marianetti,
\textit{Electronic structure calculations with dynamical mean-field theory},
Rev.\ Mod.\ Phys.\ \textbf{78}, 865 (2006).
\bibitem{Fernando2009PRB}
G.~W.~Fernando, K.~Palandage, A.~N.~Kocharian, and J.~W.~Davenport,
\textit{Pairing in bipartite and non-bipartite repulsive Hubbard clusters: octahedron},
Phys.\ Rev.\ B \textbf{80}, 014525 (2009).
\bibitem{Kocharian2009PLA}
A.~N.~Kocharian, G.~W.~Fernando, K.~Palandage, and J.~W.~Davenport,
\textit{Electron coherent and incoherent pairing instabilities in inhomogeneous bipartite and nonbipartite nanoclusters},
Phys.\ Lett.\ A \textbf{373}, 1074--1082 (2009).
\bibitem{Kocharian2008PRB}
A.~N.~Kocharian, G.~W.~Fernando, K.~Palandage, and J.~W.~Davenport,
\textit{Coherent and incoherent pairing instabilities and spin-charge separation in bipartite and nonbipartite nanocluster: Exact results},
Phys.\ Rev.\ B \textbf{78}, 075431 (2008).
\bibitem{Fernando2007PRB}
G.~W.~Fernando, A.~N.~Kocharian, K.~Palandage, T.~Wang, and J.~W.~Davenport,
\textit{Phase separation and electron pairing in repulsive Hubbard clusters},
Phys.\ Rev.\ B \textbf{75}, 085109 (2007).
\bibitem{Kocharian2007PLA}
A.~N.~Kocharian, G.~W.~Fernando, T.~Wang, K.~Palandage, and J.~W.~Davenport,
\textit{Exact thermodynamics of pairing and charge-spin separation crossovers in small Hubbard nanoclusters},
Phys.\ Lett.\ A \textbf{364}, 57--65 (2007).
\bibitem{Palandage2007JCAMD}
K.~Palandage, G.~W.~Fernando, A.~N.~Kocharian, and J.~W.~Davenport,
\textit{An exact study of pairing fluctuations and phase diagrams in four-site Hubbard Nanoclusters},
J.\ Comput.-Aided Mater.\ Des.\ \textbf{14}, 103--108 (2007).
\bibitem{Kocharian2006PRB}
A.~N.~Kocharian, G.~W.~Fernando, K.~Palandage, and J.~W.~Davenport,
\textit{Exact study of charge-spin separation, pairing fluctuations, and pseudogaps in four-site Hubbard clusters},
Phys.\ Rev.\ B \textbf{74}, 024511 (2006).
\bibitem{Kocharian2005JMMM}
A.~N.~Kocharian, G.~W.~Fernando, K.~Palandage, and J.~W.~Davenport,
\textit{Thermodynamic properties, magnetism and Mott-Hubbard-like transitions in nanoscale clusters},
J.\ Magn.\ Magn.\ Mater.\ \textbf{300}, e585--e590 (2005).
\bibitem{Tsai2006}
W.-F.~Tsai and S.~A.~Kivelson,
\textit{Superconductivity in inhomogeneous Hubbard models},
Phys.\ Rev.\ B \textbf{73}, 214510 (2006).

\end{thebibliography}

\end{document}